\documentclass[aip, jcp, reprint, floatfix]{revtex4-1}   % Journal of Chemical Physics
\usepackage{graphicx}% Include figure files
\usepackage{dcolumn}% Align table columns on decimal point
\usepackage{bm}% bold math
\usepackage[utf8]{inputenc}
\usepackage[T1]{fontenc}
\usepackage{mathptmx}
\usepackage{etoolbox}
\usepackage{subcaption}
\usepackage[version=4]{mhchem}
\usepackage{siunitx}
\usepackage{placeins}
\DeclareSIUnit\angstrom{\text {Å}}
\usepackage{tabularx}
\usepackage{graphicx} % Required for inserting images
\usepackage{tikz}
\usepackage{siunitx} % put this in your preamble
\usetikzlibrary{arrows.meta, positioning, shapes.geometric}
\tikzset{
  startstop/.style = {rectangle, rounded corners, minimum width=3cm, minimum height=1cm, text centered, draw=black, fill=red!30},
  process/.style   = {rectangle, minimum width=3cm, minimum height=1cm, text centered, draw=black, fill=blue!30},
  decision/.style  = {diamond, aspect=2, minimum width=3cm, minimum height=1cm, text centered, draw=black, fill=green!30},
  arrow/.style     = {thick,->,>=Stealth}
}

\usepackage{ulem}

\makeatletter
\def\@email#1#2{%
 \endgroup
 \patchcmd{\titleblock@produce}
  {\frontmatter@RRAPformat}
  {\frontmatter@RRAPformat{\produce@RRAP{*#1\href{mailto:#2}{#2}}}\frontmatter@RRAPformat}
  {}{}
}%
\makeatother

\begin{document}
\preprint{AIP/123-QED}

% \title{Toward Accurate Bond Breaking: Revising Natural Orbital Functionals in One-Electron Reduced Density Matrix Functional Theory}
\title{Towards more accurate natural orbital functional approximations: including 4-index cumulant contributions}

% \fundinginfo{NSERC, Canada Research Chairs, Digital Research Alliance of Canada}

% Force line breaks with \\
\author{Valerii Chuiko}
\author{Paul W. Ayers}%
\email{ayers@mcmaster.ca}
\affiliation{ 
Chemistry and Chemical Biology, McMaster University, Hamilton, Ontario, L8S 4L8, Canada%\\This line break forced with \textbackslash\textbackslash
}
\author{Eduard Matito}
\affiliation{Donostia International Physics Center (DIPC), 20018 Donostia, Euskadi, Spain}
\affiliation{Ikerbasque Foundation for Science, Plaza Euskadi 5, 48009 Bilbao, Euskadi, Spain}
\affiliation{Polimero eta Material Aurreratuak: Fisika, Kimika eta Teknologia, Kimika Fakultatea,Euskal Herriko Unibertsitatea EHU, 20080
531 Donostia, Euskadi, Spain}

% Include the name of the author that should appear in the running header

\date{\today}% It is always \today, today,

\begin{abstract}
Accurate modeling of bond breaking remains a central challenge for reduced density matrix functional theory (RDMFT). Although some modern functionals can yield reasonably accurate dissociation energies, they often fail to reproduce key properties of the dissociated fragments, such as a vanishing fragment population covariance (also known as the delocalization index) and the correct total spin angular momentum of each fragment (local spin).
In this work, we revisit the construction of natural orbital functionals by correcting the cumulant contribution produced by the PNOF5 functional. Our method enforces known contributions of the cumulant to local spin fragments and the delocalization index at the dissociation limit. We obtain the closest cumulant consistent with these physically motivated constraints and subsequently purify the corresponding one- and two-electron reduced density matrices by imposing the standard $P, Q, \text{and } G$ $N$-representability conditions. The resulting functional yields improved behavior in strongly correlated regimes. Benchmarking on the dissociation of the singlet states of \ce{N2}, \ce{NO+}, \ce{O2}, \ce{S2}, and \ce{CO} shows that in the dissociation regime the energies computed from the updated cumulant exactly reproduce the complete active space self-consistent field (CASSCF) energies. We further analyze the limitations of the approach and identify scenarios in which the current approach performs poorly. This work provides a pathway for systematically improving natural orbital functionals to achieve reliable bond-breaking calculations within RDMFT.

% Please include a maximum of seven keywords
\end{abstract}

\maketitle

\section{Introduction}
Reduced density matrix functional theory (RDMFT) provides a powerful alternative to traditional wavefunction-based electronic structure methods by shifting the focus from the exponentially scaling many-electron wavefunction to reduced density matrices (RDMs). In principle, the exact 2-electron reduced density matrix (2-RDM) contains all information needed to compute the ground-state electronic energy and other observables associated with 1- and 2-electron operators, but determining a physically valid 2-RDM requires satisfying the full hierarchy of ensemble N-representability conditions,\cite{coleman:63rmp,nakataSizeconsistencyReduceddensitymatrixMethod2012} which is computationally challenging for even the simplest systems.\cite{Alcoba2018Variational,verstichelPrimaldualSemidefiniteProgramming2011,verstichelVariationalDensityMatrix2011, Mazziotti2023Quantum,cioslowski:15jcp,ramos-cordoba:15jcp,rodriguez:17pccp,eugenedeprinceiiiVariationalDeterminationTwoelectron2024,fukudaLargescaleSemidefinitePrograms2007,nakataSizeconsistencyReduceddensitymatrixMethod2012,vanaggelenIncorrectDiatomicDissociation2009,
verstichelSubsystemConstraintsVariational2010a,
nakataVariationalCalculationsFermion2001,mazziottiQuantumChemistryWave2006,
mazziottiTwoelectronReducedDensity2012}

Instead of optimizing the 2-RDM directly, one can construct approximations for it based on simpler quantities, like the electron density and the one-electron reduced density matrix (1-RDM) for which the ensemble-N-representability problem is absent.\cite{tealeDFTExchangeSharing2022,
parrDensityFunctionalTheoryAtoms1989,
gilbertHohenbergKohnTheoremNonlocal1975,
levyUniversalFunctionalsDensity1980,
levyUniversalVariationalFunctionals1979,
valoneConsequencesExtendingMatrix1980,ayersGeneralizationsHohenbergKohnTheorem2006,
ayersGeneralizedDensityfunctionalTheory2005,
pirisNaturalOrbitalFunctional2007} These reconstructions of the 2-RDM can have lower computational cost and, if they respect some ensemble-N-representability conditions, they can be quite accurate for predicting molecular energies. 
Current methods for parameterizing the 2-RDM face two major challenges: they cannot guarantee ensemble N-representability and they focus on the electronic energy, potentially compromising the description of other molecular properties. While recent studies have explored charge localization and delocalization errors within 1-RDM and 2-RDM frameworks,\cite{Hellgren2019, LewYee2022, HuanLewYee2023} many existing functionals still lack a correct description of fragment-specific properties, including the fragment spin and the delocalization index, particularly in the dissociation limit. To develop better approximation methods, we need new physical constraints for the 2-RDM that go beyond the electronic energy.\cite{rodriguez:17pccp} 

One promising approach, the PNOFi functionals,\cite{PNOF1, Piris_helium, PNOF3, PNOF4, PNOF5, piris:14jcp,piris:17prl} use occupation number vectors to reconstruct the 2-RDM. While earlier PNOF variants focused on modeling electron pairs as quasi-independent entities, capturing both intra-pair static correlation and (optionally) inter-pair dynamic correlation\cite{Piris2013The, Piris2014Perspective,pirisExploringPotentialNatural2024}, the family has expanded to include more sophisticated treatments of electronic correlation. For instance, PNOF7\cite{PNOF7}, the Global Natural Orbital Functional (GNOF)\cite{GNOF}, and GNOFm\cite{GNOFm} significantly improve upon the independent pair approximation by explicitly incorporating inter-pair interactions, even though these improvements no longer have a simple wavefunction representation and, indeed, the refined functionals are probably not $N$-representable \cite{Erhard2026-dp}.

In this work, we focus on the PNOF5 functional, which can be represented as an antisymmetrized product of strongly orthogonal geminals (APSG),\cite{hurleyMolecularOrbitalTheory1953, parksTheorySeparatedElectron1958, parrGeneralizedAntisymmetricProduct1956,PERNAL2013127} ensuring both pure and ensemble-state N-representability. However, like APSG, it does not provide a qualitatively correct description of chemical bond breaking.\cite{Piris2014Perspective}. Despite the availability of more advanced functionals --- like PNOF7\cite{PNOF7}, GNOF\cite{GNOF}, GNOFm\cite{GNOFm} --- PNOF5 serves as an ideal benchmark for our study because its formal structure allows for a clear investigation of how new physical constraints impact the 2-RDM reconstruction.

One way to improve the performance of PNOF5 functional is to add tighter constraints on the 2-body cumulant---the correlation part of the 2-RDM.\cite{kuboGeneralizedCumulantExpansion1962,kutzelniggNormalOrderExtended1997,kutzelnigg:99jcp,mazziottiApproximateSolutionElectron1998,
ziescheCumulantExpansionsReduced2000,
ziescheRelationsCorrelationFluctuation2000} A promising approach\cite{cumulant_constraints} involves enforcing spin constraints: at the dissociation limit, the local spin\cite{ramos:12jctc} of a subsystem should match the spin of the isolated fragments. The implications of this constraint extend beyond molecular dissociation especially when describing molecular interactions in magnetic fields. The absence of systematic inclusion of \textit{local} spin conditions within existing methods causes energy levels and their degeneracy to be misrepresented in ways that compromise models for magnetic properties, e.g., associated with  electron paramagnetic resonance spectroscopy. 
Studies on hydrogen molecules in strong magnetic fields reveal that while significant interactions occur between electronic spins and magnetic fields, traditional models under 2-RDM frameworks do not accurately account for these interactions.\cite{Beau_2010, PhysRevA.53.152}
This condition is particularly relevant as it enforces conditions on the non-seniority-zero (i.e., 4-index) terms of the cumulant matrices.\cite{cumulant_constraints}

To the best of our knowledge, existing attempts to construct 2-RDM approximations from natural orbital occupations implicitly assume a seniority-zero structure to the 2-RDM, which has the advantage that the 2-RDM can be fully specified using only two indices and the disadvantage that many-electron dynamic correlation is neglected.\cite{weinholdReducedDensityMatrices1967,
    kollmarJKonlyApproximationDensity2004,
    kollmarSizeExtensiveEnergy2006,
    kollmarStructureSecondorderReduced2004,
    poelmansVariationalOptimizationSecondOrder2015a,
    head-marsdenPair2electronReduced2017,
    alcobaUnrestrictedTreatmentDirect2019,
    massaccesiVariationalDeterminationTwoparticle2021,
    onaVariationalDeterminationTwoelectron2020,
    calero-osorioSeniorityzeroWavefunctionParameterizations2025} 
This restrictions leads to the family of JKL functionals. One reason for this limitation is the paucity of practical necessary conditions for the non-seniority-zero elements of the cumulant. In this work, we employ recently developed constraints based on the local spin of fragments upon dissociation\cite{cumulant_constraints} to go beyond the JKL functional form and incorporate non-seniority-zero elements of the cumulant matrix into PNOF5 (APSG).
While enforcing such constraints may violate N-representability, we propose a method to maintain both spin constraints and necessary ensemble N-representability conditions. We believe this strategy can be further extended to other 2-RDM approximations, paving the way for the construction of more robust and accurate methods with improved energetics and molecular properties.

\section{Methodology}
% In this section, we introduce our notation and present an analytical solution to a constrained minimization problem to find the optimal cumulant tensor $\Gamma$ that minimizes deviation from a reference tensor $\Gamma_0$ while satisfying physical constraints imposed by the system's (anti)symmetry.

\subsection{Notation and background}
Following the notation in ref.~\citenum{cumulant_constraints}, we define a spinless cumulant for a singlet state in the molecular orbital basis as:
$$
\Gamma_{i j ; k l}={ }^2 D_{i j ; k l}-D_{i k} D_{j l}+\frac{1}{2} D_{i l} D_{j k},
$$
where ${ }^2 D_{i j ; k l}$ is the 2-RDM normalized to $N(N-1)$, $D_{i j}$ is the 1-RDM normalized to $N$, and $N$ is the number of electrons. 

We consider the dissociation of a molecule into two 
%identical 
fragments, $A$ and $B$. In the dissociation limit, the behavior of local spin fragments, atom-condensed cumulant contributions, and the delocalization index (DI) depend on the number $n$ of broken pairs of electrons (see Table \ref{table:PNOF5_general}).\cite{cumulant_constraints}

\renewcommand{\arraystretch}{1.5}
\begin{table}[h!]
\caption{Asymptotic values of the $\left\langle\widehat{S}^2\right\rangle_A$, atom-condensed cumulant contributions, and the delocalization index (DI) as a function of the number of broken pairs ($n$) upon dissociation.\cite{cumulant_constraints}}
\label{table:PNOF5_general}
\centering
\begin{tabular}{|l|c|c|c|c|c|c|}
\hline & $\boldsymbol{u}_{\boldsymbol{A}}$ & $\boldsymbol{\Lambda}_{\boldsymbol{A} \boldsymbol{A}}$ & $\Lambda^{\prime}{ }_{A A}$ & $\boldsymbol{\Lambda}_{\boldsymbol{A} \boldsymbol{B}}$ & $\left\langle\widehat{S}^2\right\rangle_A$ & DI \\
\hline CASSCF (2n, 2n) & $n$ & $-\frac{n}{4}$ & $\frac{n^2}{4}$ & 0 & $\frac{n}{2}\left(\frac{n}{2}+1\right)$ & 0 \\
\hline PNOF5 & $n$ & $-\frac{n}{4}$ & $\frac{n}{4}$ & 0 & $\frac{3}{4} n$ & 0 \\
\hline
\end{tabular}
\end{table}

In Table \ref{table:PNOF5_general},
\begin{equation}\label{cumulant_equations_atcond}
\begin{split}
\left\langle\widehat{S}^2\right\rangle_A &=\frac{3}
{4} u_A+\Lambda_{A A}+\Lambda_{A A}^{\prime} \\
u_A &=2 \operatorname{Tr}\left(D \mathbf{S}^A\right)-\operatorname{Tr}\left(D \mathbf{S}^A D\right) \\
\Lambda_{A A} &=\frac{1}{2} \sum_{i j k l} \Gamma_{i j ; k l} S_{k i}^A S_{l j}^A  \\
\Lambda_{A A}^{\prime} &=-\frac{1}{2} \sum_{i j k l} \Gamma_{i j ; k l} S_{l i}^A S_{k j}^A \\
\Lambda_{A B} &=\frac{1}{2} \sum_{i j k l} \Gamma_{i j ; k l} S_{k i}^A S_{l j}^B, 
\end{split}
\end{equation}
where $S_{ij}(A)$ are the elements of the overlap matrix of a fragment $A$, $\mathbf{S}^A$. The covariance population of fragments $A$ and $B$ (commonly referred as
the delocalization index (DI))\cite{matito:07fd} is defined as:\cite{mayerBondOrdersValences1986,
mayerChargeBondOrder1983,
surjanInteractionChemicalBonds1985,fradera:99jpca} 
$$
\delta(A, B)=\operatorname{Tr}\left(D \mathbf{S}^A D \mathbf{S}^B\right)-4 \Lambda_{A B}.
$$

In \cite{cumulant_constraints} authors introduced two constraints on non interacting fragments: i) the DI between two non-interacting fragments should vanish in the dissociation limit, i.e.,
$\lim _{\left|R_{A B}\right| \rightarrow \infty} \delta(A, B)=0$
and ii) the local spin of each fragment at the dissociation limit coincides with the corresponding $\left\langle\widehat{S}^2\right\rangle$ value of the isolated fragment $A$:
$\lim _{\left|R_{A B}\right| \rightarrow \infty}\left\langle\widehat{S}^2\right\rangle_A=\left\langle\widehat{S}^2\right\rangle_{\text {free A }}$

Following these results, it is shown\cite{cumulant_constraints}  different elements of the cumulant contribute unequally to $\Lambda_{AA}$ and $\Lambda_{AA}^{\prime}$. Table \ref{table:contributions_partial}\cite{cumulant_constraints} shows the contributions of selected cumulant terms obtained from CASSCF and PNOF5 calculations, whereas Table~\ref{table:contributions_partial_prime}\cite{cumulant_constraints}
presents the cumulant contributions that are nonzero for CASSCF but vanish in the case of PNOF5. In these tables, we denote the indices of ``perfect'' pairs of orbitals that share one electron with $i$ and $\bar{i}$,
so that $n_i + n_{\bar{i}} = 1$.

\begin{table}[h!]
\centering
\caption{Nonzero cumulant contributions\cite{cumulant_constraints} to $\Lambda_{AA}$ included in PNOF5 expression at the dissociation limit. The table collects the partial value of each term (multiplied by the corresponding overlaps) as a function of the number of broken pairs, $n$. There are 2$n$ terms for each contribution listed 
in this table.}
\label{table:contributions_partial}
\begin{tabular}{|l|c|c|}
\hline
 & \multicolumn{2}{c|}{Contributions to $\Lambda_{AA}$} \\
\hline
Kind of term & CASSCF & PNOF5 \\
\hline
$\Gamma_{ii;ii},\, \Gamma_{i\bar{i};\bar{i}i}$ 
    & $\frac{1}{16 n}$ 
    & $\frac{1}{16}$ \\
\hline
$\Gamma_{i\bar{i};i\bar{i}},\, \Gamma_{ii;\bar{i}\bar{i}}$ 
    & $-\frac{n+1}{16 n}$ 
    & $-\frac{1}{8}$ \\
\hline
\end{tabular}
\end{table}

\begin{table}[h!]
\centering
\caption{Nonzero cumulant contributions\cite{cumulant_constraints} to $\Lambda_{AA}^{\prime}$ not included in PNOF5 expression at the dissociation limit. The table collects the partial value of each term (multiplied by the corresponding overlaps) as a function of the number of broken pairs, $n$.}
\label{table:contributions_partial_prime}
\begin{tabular}{|l|c|c|c|}
\hline Kind of term & Number of terms & Partial & Total \\
\hline $\Gamma_{ij;ji}$ & $4 n(n-1)$ & $-\frac{1}{16 n}$ & $-\frac{n-1}{4}$ \\
\hline $\Gamma_{i \bar{j} ; j \bar{i}}, \Gamma_{i j ; \bar{j} \bar{i}}$ & $4 n(n-1) / 2$ & $\frac{n+1}{16 n}$ & $\frac{n^2-1}{4}$ \\
\hline
\end{tabular}
\end{table}

\subsection{Problem formulation}
We aim to minimize the Frobenius distance between the target cumulant $\Gamma$ and the initial cumulant $\Gamma_0$ obtained from a PNOF5 calculations,
$$
\Phi(\Gamma) = \frac{1}{2}\|\Gamma - \Gamma_0\|_F^2 = \frac{1}{2}\langle \Gamma - \Gamma_0, \Gamma - \Gamma_0 \rangle,
$$
subject to the set of linear constraints provided in Tables \ref{table:contributions_partial} and \ref{table:contributions_partial_prime},
\begin{equation}
    \frac{1}{2}\langle \Gamma, C_p^{AB} \rangle  = \lambda_p,
    \label{constraints}
\end{equation}
where the constraint tensors $C_{p;ijkl}^{AB}$ are defined by the appropriate overlap matrix elements and $\lambda_p$ represents the target constraint value.
Above we used the notation $\langle C, D\rangle \equiv \sum_{ijkl} C_{ijkl}^* D_{ijkl}$.
Specifically, referring back to Eq. \ref{cumulant_equations_atcond}, the unmasked components of the constraint tensor take the form $S_{k i}^A S_{l j}^B$ (or $-S_{l i}^A S_{k j}^B$ ), where $A$ and $B$ may refer to the same or different atoms. To ensure the tensor only selects the components that contribute to the desired value $\lambda_p$, all other elements are masked to zero. We express this by introducing a binary mask tensor $M_{p ; i j k l} \in\{0,1\}$, such that:
$$
C_{p ; i j k l}^{A B} \equiv M_{p ; i j k l}\left(S_{k i}^A S_{l j}^B\right)
$$

The mask $M_{p ; i j k l}$ is equal to 1 only for the specific index combinations $(i, j, k, l)$ that define the $p$-th constraint, and 0 otherwise.

For example, for \ce{H2} in a minimal basis set, where indices $i, j, k, l \in\{0,1\}$, the constraints are:
$\begin{aligned} 
\frac{1}{2}\left\langle\Gamma, C_1\right\rangle & =\frac{1}{2} \sum_{i, j, k, l \in\{0, 1\}} \Gamma_{i j k l}^*\left[M_{1 ; i j k l}\left(S_{k i}^A S_{l j}^B\right)\right] \\ & =\frac{1}{2}\left(\Gamma_{0000}^*\left[1 \cdot\left(S_{00}^A S_{00}^B\right)\right]+\Gamma_{0001}^*\left[0 \cdot\left(S_{00}^A S_{01}^B\right)\right]+\cdots \right) \\ 
& =\frac{1}{2} \Gamma_{0000}^* S_{00}^A S_{00}^B \\
& =\frac{1}{16}
\end{aligned}$

In the same way:
$$
\begin{aligned} 
& \frac{1}{2}\left\langle\Gamma, C_2\right\rangle=\frac{1}{2} \Gamma_{1111}^* S_{11}^A S_{11}^B=\frac{1}{16} \\ 
&\frac{1}{2}\left\langle\Gamma, C_3\right\rangle=\frac{1}{2} \Gamma_{0110}^* S_{10}^A S_{01}^B=\frac{1}{16}\\
&\frac{1}{2}\left\langle\Gamma, C_4\right\rangle=\frac{1}{2} \Gamma_{1001}^* S_{01}^A S_{10}^B=\frac{1}{16}\\
& \frac{1}{2}\left\langle\Gamma, C_5\right\rangle=\frac{1}{2} \Gamma_{0101}^*S_{00}^A S_{11}^B=-\frac{1}{8} \\ 
& \frac{1}{2}\left\langle\Gamma, C_6\right\rangle=\frac{1}{2} \Gamma_{1010}^*S_{11}^A S_{00}^B=-\frac{1}{8} \\ 
& \frac{1}{2}\left\langle\Gamma, C_7\right\rangle=\frac{1}{2} \Gamma_{0011}^* S_{10}^A S_{10}^B=-\frac{1}{8}\\
& \frac{1}{2}\left\langle\Gamma, C_8\right\rangle=\frac{1}{2} \Gamma_{1100}^* S_{01}^A S_{01}^B=-\frac{1}{8}
\end{aligned}
$$

This constrained optimization problem can be solved, analytically, using Lagrange multipliers. Specifically, the Lagrangian is
\begin{equation}
\begin{split}
\mathcal{L}_p(\Gamma, \eta_p) =& \frac{1}{2}\langle \Gamma - \Gamma_0, \Gamma - \Gamma_0 \rangle\\ &
- \eta_p\left(\frac{1}{2}\langle C_p^{AB}, \Gamma \rangle - \lambda_p\right),
\end{split}
\end{equation}
where $\eta_p$ is the Lagrange multiplier enforcing the constraint. Because different constraints affect disjoint index sets of the cumulant, imposing the constraints sequentially and simultaneously is equivalent.

After taking partial derivatives and solving the system of linear equations, we obtain
$$
\Gamma = \Gamma_0 + \frac{2\lambda_p - \langle C_p^{AB}, \Gamma_0 \rangle}{\langle C_p^{AB}, C_p^{AB} \rangle} C_p^{AB}
$$
However, this procedure does not guarantee the ensemble N-representability of the 2-RDM corresponding to the updated cumulant. We therefore project the updated 2-RDM onto the set of 2-RDMs that satisfy some necessary ensemble N-representability conditions.  To compute the closest positive semidefinite matrix satisfying the desired constraints, we used the algorithm introduced by Lanssens et al. \cite{closest_Nrep} The closest symmetric positive semidefinite matrix $B$ to a matrix $B_0$ with fixed trace $T$ is obtained by shifting the eigenvalues $\{\lambda_i\}$ of $B_0$ by a value $\sigma_0$, which is the root of the equation $f(\sigma) = T$, where $f(\sigma) = \sum_i \theta\left(\lambda_i-\sigma\right)\left(\lambda_i-\sigma\right)$ and $\theta\left(\lambda_i-\sigma\right)$ is the Heaviside step function. In this work, we used only the P, Q, and G conditions since, as shown in section \ref{results}, imposing the subsystem constraints on the cumulant does not lead to large violations of the ensemble N-representability constraints.

Since the original constraints\cite{cumulant_constraints} on the cumulants are derived in spatial basis and the P, Q, and G conditions are derived in spin-orbital basis, it is necessary to map the 2-RDM that corresponds to the corrected cumulant from the spatial basis to the spin-orbital basis before projecting onto the selected ensemble N-representability constraints. As derived in the supplementary material, we compute the $\alpha\alpha$, $\alpha\beta$ and $\beta\beta$ components of 2-RDM, $D^{\sigma_1\sigma_2\sigma_3\sigma_4}_{pqrs}$, using:

\begin{equation}
\label{spatial2spin}
\begin{split}
&  D_{i j k l}^{\alpha \alpha \alpha \alpha}=A_{i j k l}-A_{i j l k} \\
&  D_{i j k l}^{\beta \beta \beta \beta}=A_{i j k l}-A_{i j l k} \\
&  D_{i j k l}^{\alpha \beta \alpha \beta}=A_{i j k l} \\
&  D_{i j k l}^{\beta \alpha \beta \alpha}=A_{i j k l} \\
&  D_{i j k l}^{\alpha \beta \beta \alpha}=-A_{i j l k} \\
&  D_{i j k l}^{\beta \alpha \alpha \beta}=-A_{i j l k},
\end{split}
\end{equation}

where $D_{pqrs}$ is the spatial 2-RDM and $A_{i j k l}=\frac{2 {D_{i j k l}}+D_{i j l k}}{6}$. 

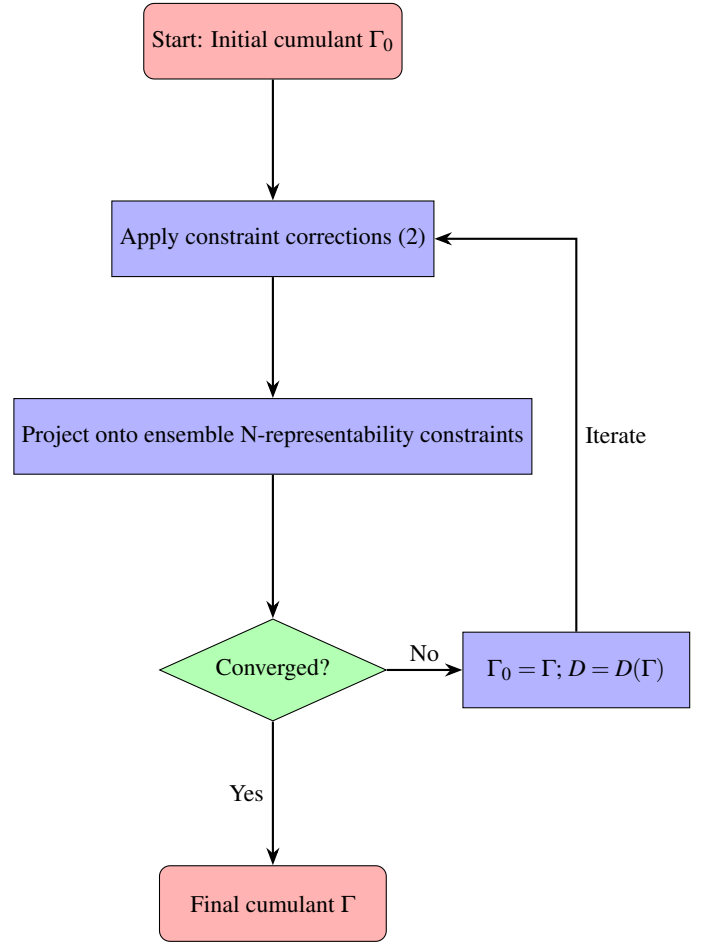
\begin{figure}[ht]
  \centering
  \begin{tikzpicture}[node distance=1.6cm and 1.0cm, auto]

    % Nodes
    \node (start)    [startstop]                         {Start: Initial cumulant $\Gamma_0$};
    \node (process1) [process, below=of start]           {Apply constraint corrections (\ref{constraints})};
    \node (process3) [process, below=of process1]        {Project onto ensemble N-representability constraints};
    \node (decision) [decision, below=of process3, yshift=-3mm] {Converged?};
    \node (process4) [process, right=of decision]        {$\Gamma_0 = \Gamma$; $D = D(\Gamma)$};
    \node (end)      [startstop, below=of decision, yshift=-3mm] {Final cumulant $\Gamma$};

    % Edges
    \draw [arrow] (start)    -- (process1);
    \draw [arrow] (process1) -- (process3);
    \draw [arrow] (process3) -- (decision);

    % Straight labelled arrow from decision to the right (No)
    \draw [arrow] (decision.east) -- node[midway, above]{No} (process4.west);

    % Return arrow from process4 back to process1 (loop)
    \draw [arrow] (process4.north) |- (process1.east) node[pos=0.25, right]{Iterate};

    % Yes path downwards from decision to end
    \draw [arrow] (decision) -- node[midway, left]{Yes} (end);

  \end{tikzpicture}
  \caption{Schematic iterative procedure: impose the subsystem constraints on the cumulant, project the 2-RDM $D$ onto the (approximate) set of ensemble $N$-representable 2-RDMs using the algorithm from reference \citenum{closest_Nrep}; repeat until convergence.}
  \label{fig:iterative-loop}
\end{figure}

Note that the resulting projected cumulant may violate the originally imposed constraints in eq.~\ref{constraints}. Therefore, we need to iteratively reimpose the constraints on the projected cumulant, and then reimpose the ensemble $N$-representability conditions on the constrained cumulant, as shown in {Fig.}~\ref{fig:iterative-loop}. The convergence criteria is considered achieved when all the constrained reached an error $\le 10^{-5}$ or when all all constraints has not changed for 10 iterations.

\section{Tested systems}
We tested this method for diatomic molecules because this avoids ambiguity about how to define fragments. Specifically, we considered singlet states of \ce{N2}, \ce{NO+}, \ce{O2}, \ce{S2}, and \ce{CO}. For each system, the cumulant $\Gamma_0$ was obtained from PNOF5 functional using the DoNOF software\cite{Piris2021} using integrals computed from the STO-3G basis set.\cite{hehreSelfConsistentMolecularOrbitalMethods1969,sun2018pyscf,kimGBasisPythonLibrary2024} All systems in this study were treated as singlets, since the cumulant constraints at the dissociation limit \cite{cumulant_constraints} have been developed exclusively for singlet states. In all calculations, we set the parameter corresponding to the maximum index of natural orbitals with occupation numbers equal to 1.0 to four, as this was found to improve orbital convergence. Furthermore, along the dissociation curve, the orbitals obtained at each geometry were used as the initial guess for the subsequent step, ensuring a consistent and stable optimization procedure.

For \ce{N2}, \ce{NO+}, \ce{O2}, and \ce{S2}, we used the algorithm summarized in Figure~\ref{fig:iterative-loop}. 
However, for the \ce{CO} molecule we introduced some modifications because, upon dissociation, the asymptotic values reported in Tables 1 and 2 are correct only when the subspace overlap matrix satisfies $S^A_{i i} = 1/2$ for all pairs of orbitals.\cite{cumulant_constraints} This condition corresponds to an equal partitioning of the bonding orbitals between the two asymptotically separated fragments. However, in an unrestricted or symmetry-broken dissociation of neutral \ce{CO} the $\sigma$ orbitals naturally localize on the individual atoms at large internuclear separation, causing the corresponding elements of $\mathbf{S}^A$ to approach 1 or 0. To recover the desired asymptotic behavior, we instead enforce an ionic dissociation limit (C$^-$,  O$^+$), which preserves the delocalized character of the $\sigma$ orbital. 

This is achieved by modifying the diagonal elements of one-electron integrals: small potentials are added to the diagonal elements corresponding to the carbon atomic orbitals and subtracted from those corresponding to the oxygen orbitals. The absolute value of the \textit{potential on each atom} is computed as $\text{EA}_C \left(1-e^{R/2}\right)$, where $\text{EA}_C$ is the electron affinity of carbon atom and $R$ is the interatomic distance.
These adjustments shift the relative orbital energies, favoring electron localization on carbon and depletion on oxygen, thereby stabilizing a $\sigma$ orbital that is equally shared across the fragments and producing the $\mathbf{S}^A_{ii} = 1/2$ values consistent with the asymptotic limits reported in Tables~\ref{table:contributions_partial} and \ref{table:contributions_partial_prime}. Importantly, although this condition was not explicitly enforced, it is also satisfied for \ce{NO+}, resulting in a final charge distribution consistent with correct dissociation limit, \ce{N\bond{...}O+}.

\section{Results and discussions}\label{results}
In this section, we present results for \ce{N2}, \ce{O2}, and \ce{NO+} in their dissociation regime. Results for other systems are qualitatively similar and reported in the supplementary materials. We focus on these systems because they pertain to especially difficult systems. \ce{N2} is difficult to treat at the seniority-zero level and is thus expected to be especially challenging for a JKL functional, suggesting that 4-index components of the cumulant may be especially important.\cite{bytautasSeniorityNumberDescription2015,
calero-osorioSeniorityZeroCanonicalTransformation2026,
vanaggelenChemicalVerificationVariational2010a} Singlet \ce{O2} is an excited state, allowing us to test whether our formulation works for the lowest excited state of a given spin symmetry. \ce{NO} is a notorious example where variational 2-electron reduced density matrix methods give fractional charges at dissociation, providing a stringent test of our $N$-representability constraints.\cite{vanaggelenIncorrectDiatomicDissociation2009}

\begin{figure}
    \centering
    \includegraphics[width=0.9\linewidth]{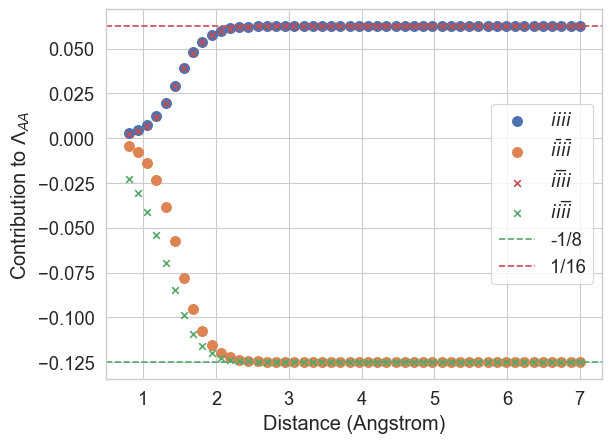}
    \caption{Contributions of selected elements of the cumulant during dissociation of N$_2$; $i=1,\bar{i}=4$}
    \label{fig:N2_contributions}
\end{figure}

\begin{figure}
    \centering
    \includegraphics[width=0.9\linewidth]{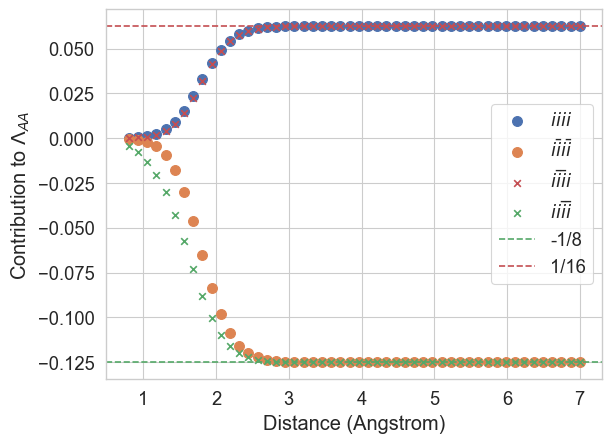}
    \caption{Contributions of selected elements of the cumulant during dissociation of O$_2$; $i=0,\bar{i}=3$}
    \label{fig:O2_contributions}
\end{figure}

\begin{figure}
    \centering
    \includegraphics[width=0.9\linewidth]{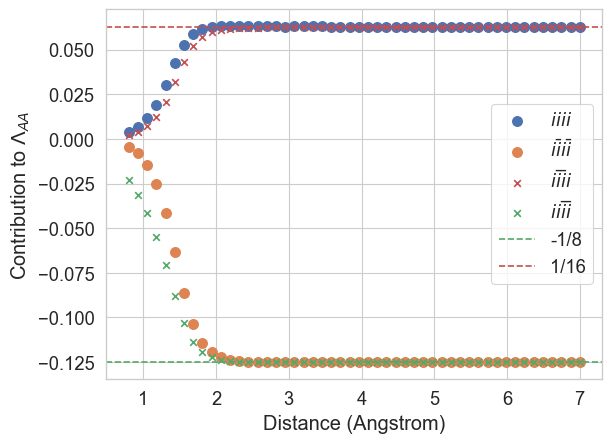}
    \caption{Contributions of selected elements of the cumulant during dissociation of NO$^+$; $i=1,\bar{i}=4$}
    \label{fig:NO_contributions}
\end{figure}

To verify the calculated reduced density matrices, we examined the asymptotic behavior of certain cumulant elements and compared them to those in ref.~\citenum{cumulant_constraints}. The asymptotic values are shown in Figures~\ref{fig:N2_contributions},  \ref{fig:O2_contributions}, and \ref{fig:NO_contributions}. 

\begin{figure}
    \centering
    \includegraphics[width=0.9\linewidth]{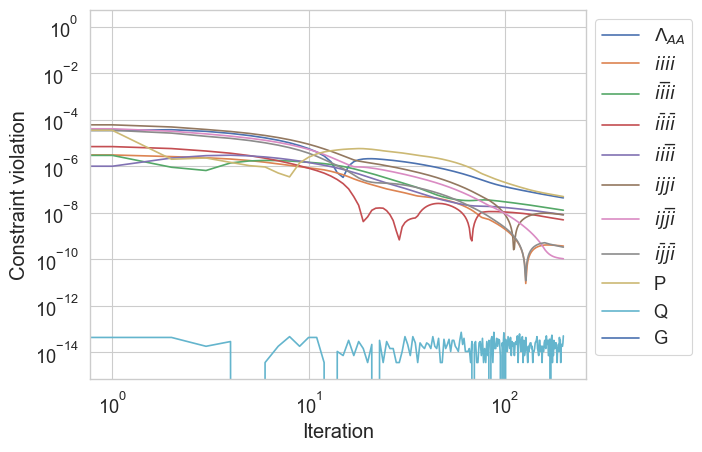}
    \caption{Change of violations of constraints of N$_2$ molecule at dissociation regime during iterative procedure.}
    \label{fig:N2_iterations}
\end{figure}

\begin{figure}
    \centering
    \includegraphics[width=0.9\linewidth]{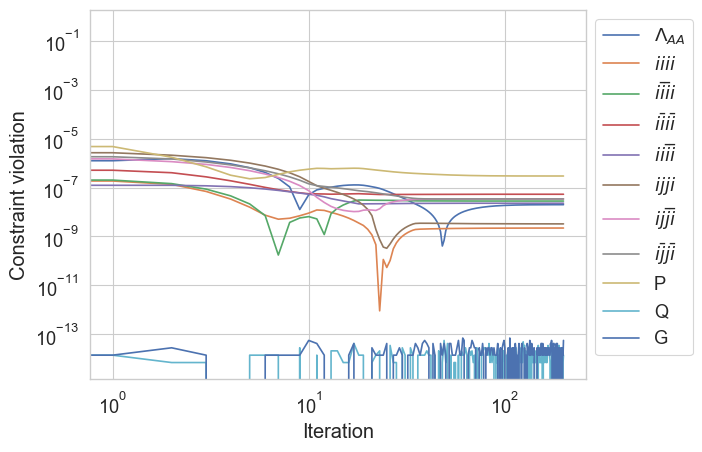}
    \caption{Change of violations of constraints of O$_2$ molecule at dissociation regime during iterative procedure.}
    \label{fig:O2_iterations}
\end{figure}

\begin{figure}
    \centering
    \includegraphics[width=0.9\linewidth]{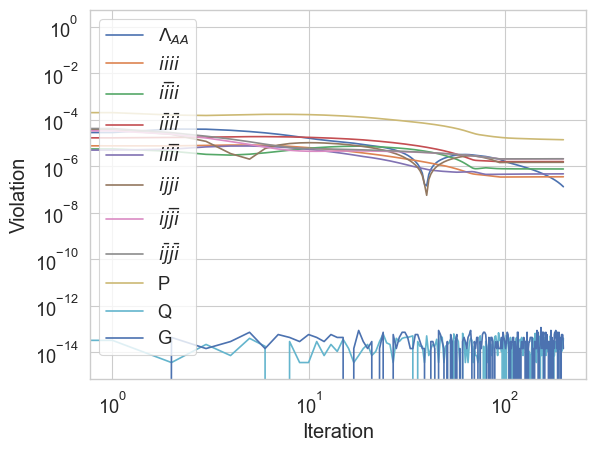}
    \caption{Change of violations of constraints of NO$^+$ molecule at dissociation regime during iterative procedure.}
    \label{fig:NO_iterations}
\end{figure}

For geometries in the dissociation regime ($\ge 3.5$ \si{\angstrom{}}), we applied the algorithm shown in Figure 1. To study changes in element contributions and violations of N-representability constraints, we ran the algorithm for 200 iterations for a \ce{N2} at an internuclear distance of 4.98 \si{\angstrom{}}. Figures \ref{fig:N2_iterations}, \ref{fig:O2_iterations} and \ref{fig:NO_iterations} show violations of cumulant element contributions and N-representability constraints. In all cases, the lowest eigenvalues of the $P, Q$ and $G$ matrices were $\gtrsim - 10^{-5}$, establishing that the $N$-representability constraints are satisfied.

Figure~\ref{fig:N2_iterations} shows that all constraints converged during optimization. Notably, even after the first iteration, the P condition violation is on the order of $10^{-6}$, while the Q-condition is fully satisfied. These results suggest that imposing these constraints on the cumulant does not violate the ensemble N-representability of the corresponding RDM, therefore, there is scope for introducing 4-index corrections to existing JKL (2-index) RDMFT functional approximations.

Figure~\ref{fig:O2_iterations} shows the change in constraint violations for \ce{O2} at the same interatomic distance. All constraints converged similarly. However, violations of the P-condition and the cumulative contribution of $ii\bar{i}\bar{i}$ elements of the cumulant reached plateaus at $\sim10^{-6}$ and $\sim10^{-7}$, respectively. A similar trend is observed for \ce{NO+}, where the P-condition violation plateaus at $\sim10^{-5}$, and the contribution of the $ii\bar{i}\bar{i}$ elements of the cumulant stops improving after reaching an error of $\sim10^{-6}$. This behavior may be due to (a) the molecules not being sufficiently close to the dissociation limit and (b) residual errors from the orbital optimization of the PNOF5 computation.

Tables \ref{tab: O2 compacted forms} and \ref{tab: N2 NO+ compacted forms} present the asymptotic values of $\left\langle \widehat{S}^2 \right\rangle_A$, $\Lambda$, the compacted forms, and the delocalization index (DI) upon dissociation of the \ce{O2}, \ce{N2}, and \ce{NO+} molecules. As expected, after correcting the contributions of the non-seniority-zero cumulant terms to $\Lambda_{AA}$ and $\Lambda^{\prime}_{AA}$, the resulting values of $\left\langle \widehat{S}^2 \right\rangle_A$ and $\Lambda^{\prime}_{AA}$ exactly match the theoretical CAS-SCF limits. Notably, quantities that already exhibit correct asymptotic behavior at the PNOF5 level ($u_A$, DI, $\Lambda_{AA}$, and $\Lambda_{AB}$) remain unchanged.

% EDU: round up to three digits to fit the size of the column. we could add more digits but I belive three is enough for our purposes.
\sisetup{
  round-mode = places,
  round-precision = 3
}

\begin{table}[h!]
\centering
\caption{Asymptotic values of the $\left\langle\widehat{S}^2\right\rangle_A$, compacted forms, and the DI upon dissociation of the O$_2$ molecule.}
\label{tab: O2 compacted forms}
\begin{tabular*}{\linewidth}{|@{\extracolsep{\fill}}|l|S|S|S|S|S|S|}
\hline
 & {${u}_{{A}}$} 
 & {${\Lambda}_{{A}{A}}$} 
 & {$\Lambda_{AA}'$} 
 & {$\Lambda_{AB}$} 
 & {$\langle \widehat{S}^2 \rangle_A$} 
 & {DI} \\
\hline
CASSCF(4; 4)
 & 2.0000000000023057
 & -0.4999999999999889
 & 0.9999999999952209
 & 0.
 & 1.9999999999969613
 & 0. \\
\hline
Corrected PNOF5
 & 1.9999996226832977
 & -0.4999999999999856
 & 1.0000002030284991
 & 0.0000
 & 1.9999999200409868
 & 0.0000 \\
\hline
PNOF5
 & 1.9999993072004152
 & -0.49999965320955214
 & 0.49999965205343233
 & 0.0000
 & 1.4999994792441917
 & 0.0000 \\
\hline
\end{tabular*}
\end{table}

\begin{table}[h!]
\centering
\caption{Asymptotic values of the $\left\langle\widehat{S}^2\right\rangle_A$, compacted forms, and the DI upon dissociation of the N$_2$ and NO$^+$ molecules.}
\label{tab: N2 NO+ compacted forms}
\begin{tabular*}{\linewidth}{|@{\extracolsep{\fill}}|l|S|S|S|S|S|S|}
\hline
 & {${u}_A$}
 & {${\Lambda}_{AA}$}
 & {$\Lambda_{AA}'$}
 & {$\Lambda_{AB}$}
 & {$\langle \widehat{S}^2 \rangle_A$}
 & {DI} \\
 \hline
CASSCF(6; 6)
 & 2.99999998
 & -0.749999997
 & 2.24999999
 & 0.
 & 3.74999998
 & 0. \\
\hline
Corrected PNOF5
 & 3.00000713
 & -0.75000000
 & 2.25000397
 & 0.
 & 3.75000931
 & 0. \\
\hline
PNOF5
 & 3.00000412
 & -0.75000103
 & 0.74999782
 & 0.
 & 2.24999987
 & 0. \\
\hline
\end{tabular*}
\end{table}

% \begin{table}[h!]
% \centering
% \caption{Asymptotic values of the $\left\langle\widehat{S}^2\right\rangle_A$, compacted forms, and the DI upon dissociation of the NO$^+$ molecule.}

% \begin{tabular*}{\linewidth}{|@{\extracolsep{\fill}}|l|S|S|S|S|S|S|}
% \hline
%  & {${u}_A$}
%  & {${\Lambda}_{AA}$}
%  & {$\Lambda_{AA}'$}
%  & {$\Lambda_{AB}$}
%  & {$\langle \widehat{S}^2 \rangle_A$}
%  & {DI} \\
% \hline
% CASSCF
%  & 3.00000000
%  & -0.75000000
%  & 2.25000000
%  & 0.
%  & 3.75000000
%  & 0. \\
% \hline
% Corrected PNOF5
%  & 2.99998935
%  & -0.75000000
%  & 2.25000154
%  & 0.
%  & 3.74999355
%  & 0. \\
% \hline
% PNOF5
%  & 2.99998420
%  & -0.74999578
%  & 0.74999461
%  & 0.
%  & 2.24998698
%  & 0. \\
% \hline
% \end{tabular*}
% \end{table}

Also, we compare the energies of the corrected cumulant using the proposed algorithm with energies from direct CASSCF calculations for selected geometries in the dissociation regime (See Table \ref{tab:energies_N2_O2}.). These energies are obtained by averaging the energies of the 10 most-stretched geometries. Note that the contributions of elements of the cumulant \cite{cumulant_constraints} are correct only at full dissociation. Until this point is reached, the corrected cumulant values are approximations. For incomplete dissociation, imposing these constraints can lead to energies above or below the corresponding CASSCF energies. This behavior is seen for \ce{S2} (see the supporting information), where the energy after improving the cumulant converges to the CASSCF energy from below, reaching it at interatomic distances greater than $4$ \si{\angstrom{}}.

\begin{table}[h!]
\centering
\caption{Energies (Ha) for N$_2$, O$_2$, and NO$^+$ molecules}
\label{tab:energies_N2_O2}
\begin{tabular*}{\linewidth}{|@{\extracolsep{\fill}}c|S|S|S|}
\hline
Molecule
& \multicolumn{1}{c|}{PNOF5}
& \multicolumn{1}{c|}{Corrected Energy}
& \multicolumn{1}{c|}{CASSCF (2n, 2n)} \\
\hline
N$_2$
& -107.31466291827358
& -107.43799178325571
& -107.43802430912459 \\
\hline
O$_2$
& -147.56086178133913
& -147.60830284005132
& -147.60836236196445 \\
\hline
NO$^+$
& -127.02983188486697
& -127.16268610731247
& -127.16269299987349 \\
\hline
\end{tabular*}
\end{table}

To address the practical usage of the method we need to discuss its robustness and theoretical scaling. 
The proposed method follows a two-stage approach: a foundational PNOF5 calculation followed by the iterative eigenvalue decomposition of the 2-RDM. Since the 2RDM is constructed in a spin-orbital basis of size $(2 N)^2 \times(2 N)^2$, the diagonalization step scales formally as $O\left(\left(N^2\right)^3\right)$, or $O\left(N^6\right)$, where $N$ represents the number of spatial orbitals. While $O\left(N^6\right)$ is computationally intensive, it remains strictly polynomial, offering a significant theoretical advantage over multiconfigurational alternatives.
In contrast, the CASSCF $(2 n, 2 n)$ method is governed by the "exponential wall" inherent to the Full Configuration Interaction (FCI) expansion within the active space. The number of configurations for an active space of $2 n$ electrons in $2 n$ orbitals grows factorially, making CASSCF expensive for all but the smallest systems. However, CASSCF may demonstrate superior performance over the proposed method if three specific criteria are met 1) The number of active electrons ( $n$ ) remains small, 2)The total number of spatial orbitals ( $N$ ) is large, and 3) The system is far from the dissociation limit, where the proposed method may require multiple iterations to achieve convergence.
In practice, when the method is applied at the dissociation limit ( $r \geq 5 \si{\angstrom{}}$ ), ensemble $N$-representability is naturally maintained. In this regime, the computational cost is effectively reduced to the scaling of the linear projection step, which is merely $O\left(N^4\right)$.

Finally, to verify the robustness of our approach we repeated our calculation for \ce{N2} with three more basis sets -- 6-311G, cc-pVDZ, and cc-pVTZ. In Table \ref{tab:PNOFi methods} we present energy, $\left\langle\widehat{S}^2\right\rangle_A$, and $\Lambda^{'}_{AA}$ at the dissociation limit for 6-311G basis set.  (Results for cc-pVDZ and cc-pVTZ basis sets are qualitatively the same and are present in the SI.) Values for compacted forms that are the same for all included methos are not included in the table and are omitted from all further discussion. 

\begin{table}[h!]
\centering
\caption{Asymptotic values of the energy, $\left\langle\widehat{S}^2\right\rangle_A$, and the $\Lambda^{'}_{AA}$ dissociation of the N$_2$ molecule.}

\begin{tabular*}{\linewidth}{|@{\extracolsep{\fill}}|l|S|S|S|}
\hline
 & {$E, \text{Ha}$}
 & {$\Lambda_{AA}'$}
 & {$\langle \widehat{S}^2 \rangle_A$}\\
\hline
CAS-SCF(6; 6)
& -108.78947825
& 2.250004
& 3.7499999\\
\hline
Corrected PNOF5
 & -108.78947
 & 2.24999987
 & 3.74999979\\
\hline
PNOF7
 & -108.78951
 & 0.74999998
 & 2.24999995\\
\hline
GNOF
 & -108.76292
 & 0.74997537
 & 2.24997792\\
 \hline
 PNOF5
 & -108.6834
 & 0.74999782
 & 2.24999987\\
\hline

\end{tabular*}
\label{tab:PNOFi methods}
\end{table}

The results presented in Table ~\ref{tab:PNOFi methods} show that the electronic energy obtained with PNOF7 ($E=-108.7895 \mathrm{Ha}$ ) is in excellent agreement with both the $\operatorname{CASSCF}(6,6)$ and the corrected PNOF5 results, effectively matching them to high precision. However, the values of $\left\langle\widehat{S}^2\right\rangle_A$ and $\Lambda_{A A}^{\prime}$ are identical for PNOF5, GNOF, and PNOF7, and deviate from the CAS-type limit, suggesting that PNOF7 gets ``the right answer for the wrong reason''. 

The terms $\Gamma_{i j ; j i}$, $\Gamma_{i \tilde{j} ; \tilde{j}i}$, and $ \Gamma_{i j ; \tilde{j}\tilde{i}}$ do not enter the expression of $\Lambda_{A A}$ whereas $\Lambda_{A A}^{\prime}$ does have contributions from these terms. On the other hand, the PNOF5 functional relies on a simplified cumulant structure built strictly upon the independent-pair approximation, wherein correlation is confined to isolated orbital pairs: one below the Fermi level coupled to one above it. Consequently, PNOF5 enforces vanishing off-diagonal, inter-pair cumulant terms.
Another observation is that the correct energy in the asymptotic limit is ensured by assuring that 2-RDM terms where all indices are assigned to the same fragment are correct. The inter-fragment terms are not zero---they are required by the $N$-representability constraints and the correct description of the dissociated fragments' spin---but they do not affect the asymptotic energy.

While these off-diagonal terms mathematically drop out of the intra-pair energy expression $\left(\Lambda_{A A}\right)$, they contribute to local spin properties captured by $\Lambda_{A A}^{\prime}$. As detailed in Table \ref{table:contributions_partial_prime}, the exact CASSCF wave function demonstrates that non-zero inter-pair cumulants---specifically $\Gamma_{i j ; j i}$, $\Gamma_{i j ; j i}$, and $\Gamma_{i j ; \tilde{j} \tilde{i}}$---are essential at the dissociation limit.
The physical role of these missing terms is two-fold:
\begin{enumerate}
\item Linear Spin Contributions: The $\Gamma_{i j ; j i}$ terms describe the coupling between different strongly occupied orbitals (below the Fermi level). These introduce a linear dependency in $n$ to the local spin variance $\Lambda_{A A}^{\prime}$. Interestingly, the magnitude of the $\Gamma_{i j ; j i}$ contribution to the local spin is identical to that of the intra-pair $\Gamma_{i \tilde{i} ; \tilde{i} i}$ term (which is natively included in PNOF5).
\item Quadratic Spin Dependencies: The $\Gamma_{i \tilde{j} ; \tilde{j} i}, \Gamma_{i j ; \tilde{j} \tilde{i}}$ terms describe couplings across two distinct pairs, $(i, \tilde{i})$ and $(j, \tilde{j})$. These specific inter-pair correlations are strictly responsible for recovering the $n^2$ dependence of the local spin when multiple bonds dissociate. This quadratic scaling arises from the total combinatorics of the interacting terms on the dissociated fragment, a collective phenomenon that independent-pair models inherently miss.
\end{enumerate}

%This discrepancy indicates that $N$-representability is not fully achieved in the asymptotic region for these functionals.

This leads to the conclusion that violation of constraints on $\Lambda^\prime_{AA}$ introduces spin-contamination of different fragments at the dissociation in the NOF family of functionals. 
In the NOF framework, this contamination arises from absent or poorly approximated inter-pair cumulant terms, yielding an incorrect local spin expectation value $\left(\left\langle\hat{S}_A^2\right\rangle\right)$. Our method directly resolves this artifact by mathematically constraining the compacted forms to enforce the exact asymptotic spin contributions for each fragment separately. Imposing this strict physical boundary condition on the localized density matrix prevents spin contamination, guaranteeing that the dissociation profile yields physically meaningful, pure-spin fragments at the asymptotic limit, as long as imposing constraints are reached during the iterative procedure.
Also, the total molecular $\langle \hat{S}^2 \rangle$ is strictly guaranteed in our formalism because we map the spatial 2-RDM into a spinized 2-RDM, under the assumption that system is singlet. This mapping inherently enforces a total spin of zero during the computation of the $G$-condition, ensuring that the overall singlet state of the molecule is mathematically preserved.\cite{vanaggelenConsiderationsDescribingNonsinglet2012}(Higher-spin states could also be treated, but would require a more sophisticated treatment.)

This limitation of NOF family highlights the potential for extending our approach to other functionals and additional physical constraints. Nonetheless, it is likely that exact $N$-representability cannot be simultaneously enforced with cumulant constraints \cite{cumulant_constraints} for methods like GNOF/PNFO7, suggesting that an alternative strategy may be required. Rather than jointly projecting onto selected ensemble $N$-representability constraints while simultaneously enforcing cumulant constraints, a more robust approach would be to first project the cumulant onto its corresponding constraint set only once, and then iteratively determine the closest 2-RDM satisfying the chosen $N$-representability conditions. Potential methods for this procedure include Halpern \cite{Halpern1967FixedPO} and Dykstra \cite{boyle1986method} algorithms. However, cyclic projection methods do not guarantee preservation of the correct cumulant trace in the corrected 2-RDM.
We focused on spin-fragments because spin remains a good quantum number for dissociating polyatomic fragments, but we could extend the methodology to orbital- and total-angular momentum, which are also good quantum numbers for molecular dissociation into atomic fragments. However, for processes where the dominant static correlation is not associated with spin-symmetry breaking in the dissociation limit (e.g., cases where restricted Hartree-Fock is qualitatively correct), we expect PNOF5 and its refinements to be qualitatively correct and we do not expect this method to provide a substantive improvement over the uncorrect natural orbital functional. Similarly, when there is substantial static correlation at dissociation that is not associated with spin-symmetry-breaking, we expect our approach will capture only a partial description of the static correlation at dissociation.

We leave the development and investigation of this strategy for future work.

% \ref{fig:N2_energy} and \ref{fig:O2_energy}, the energy corresponding to the corrected cumulant matches the CASSCF energy. \edu{[EDU: I would not include this information as figure, considering we are showing three straight lines.]}

% \begin{figure}[h!]
%     \centering
%     \includegraphics[width=0.9\linewidth]{figures/N2_energy.png}
%     \caption{Energies of $N_2$ molecule at dissociation computed with CASSCF(6, 6), DOCI wavefunction, and cumulant that satisfies set of constraint\cite{cumulant_constraints}} 
%     \label{fig:N2_energy}
% \end{figure}

% \begin{figure}[h!]
%     \centering
%     \includegraphics[width=0.9\linewidth]{figures/O2_energy.png}
%     \caption{Energies of $O_2$ molecule at dissociation computed with CASSCF(4, 4), DOCI wavefunction, and cumulant that satisfies set of constraint\cite{cumulant_constraints}} 
%     \label{fig:O2_energy}
% \end{figure}

\section{Summary}
This work demonstrates that enforcing the correct asymptotic behavior of fragment spin and cumulant (de)localization indices imposes stringent, structurally transparent, constraints on the two-electron cumulant.
Most importantly, the results show that enforcing physically mandated asymptotics does not require large or global modifications of the cumulant. This opens a clear conceptual path for extending this strategy to other 2-RDM approximations: first, a parametrization of the non-seniority-zero (4-index) terms of the 2-body cumulant is required. Once these terms are defined, the proposed algorithm can be exploited to incorporate the required correlation while simultaneously satisfying local spin constraints and ensemble $N$-representability conditions. Such approach allows to introduce exact asymptotic constraints directly into the cumulant parametrization, rather than relying on \textit{ad hoc} occupation-based heuristics or \textit{post hoc} regularization. The framework introduced here---combining analytic constraint enforcement with further purification via $N$-representability projections---provides a practical template for such developments.

More broadly, we suggest that successful many-electron functionals must treat the dissociation limit not as a pathological edge case but as a rigorous boundary condition that shapes the internal structure of the cumulant across the full bonding regime. Embedding these constraints into the functional form promises systematically improved behavior in bond breaking, charge localization, and spin separation, and offers a route toward new cumulant-based models that are both computationally economical and aligned with the exact theory.

\section{Supplementary Material}
In the Supplementary Materials, we present plots showing the contributions of cumulant elements during molecular dissociation, the evolution of constraint violations in the dissociation regime throughout the iterative optimization procedure, and energy profiles as functions of interatomic distance for PNOF5, the improved functional, and CASSCF calculations for the \ce{S2} and \ce{CO} molecules.

\begin{acknowledgments}
PWA and VC thank the Canada Research Chairs (CRC-2022-00196), the Digital Research Alliance of Canada, NSERC (ALLRP/592521-2023 and RGPIN-2024-06707), and MITACS for financial and computational support. EM is grateful for the funding and technical support provided by the
Donostia International Physics Center (DIPC), a Basque Government project (IT2067-26) and the grant PID2022-140666NB-C21 funded by MCIN/AEI/10.13039/501100011033 and “FEDER Una manera de hacer Europa”. The authors also thank the DoNOF development team, in particular Dr. Juan Felipe Huan Lew-Yee, for technical support. 
\end{acknowledgments}

\section*{Data Availability Statement}
The data that support the findings of this study are available from the corresponding author upon reasonable request. 

\section{References}
\bibliographystyle{aipnum4-1}
\bibliography{bibliography}

\end{document}

\preprint{AIP/123-QED}
\section{Supplementary Material}
In the Supplementary Materials, we derive the mapping of the two-electron reduced density matrix (2RDM) from spatial to spin-orbital basis for singlet states. We also present plots showing the contributions of cumulant elements during molecular dissociation, the evolution of constraint violations in the dissociation regime throughout the iterative optimization procedure, and energy profiles as functions of interatomic distance for PNOF5, the improved functional, and CASSCF calculations for the \ce{S2} and \ce{CO} molecules.
\renewcommand{\thefigure}{SE\arabic{figure}}
% \setcounter{figure}{0}
% \begin{figure}[H] % requires \usepackage{float}
%     \centering
%     \includegraphics[width=0.8\linewidth]{figures/NO_contributions.png}
%     \caption{Contributions of selected elements of the cumulant during dissociation of $NO^+$}
% \end{figure}

% \begin{figure}[H]
%     \centering
%     \includegraphics[width=0.8\linewidth]{figures/NO_iters.png}
%     \caption{Change of violations of constraints of $NO^+$ molecule at dissociation regime during iterative procedure.}
% \end{figure}

% \begin{table}[h!]
% \centering
% \caption{Asymptotic values of the $\left\langle\widehat{S}^2\right\rangle_A$, compacted forms, and the DI upon dissociation for the $NO^+$ molecule.}
% \begin{tabular}{|l|S|S|S|S|S|S|}
% \hline
%  & {${u}_A$}
%  & {${\Lambda}_{AA}$}
%  & {$\Lambda_{AA}'$}
%  & {$\Lambda_{AB}$}
%  & {$\langle \widehat{S}^2 \rangle_A$}
%  & {DI} \\
% \hline
% CASSCF
%  & 3
%  & -0.75
%  & 2.25
%  & 0.
%  & 3.75
%  & 0. \\
% \hline
% Fixed PNOF5
%  & 2.99998935
%  & -0.750000000013
%  & 2.25000154
%  & 0.00000280
%  & 3.74999355
%  & 0.00002041 \\
% \hline
% PNOF5
%  & 2.99998420
%  & -0.74999578
%  & 0.74999461
%  & -0.00000013
%  & 2.24998698
%  & 0.00001662 \\
% \hline
% \end{tabular}
% \end{table}

% \begin{figure}[H]
%     \centering
%     \includegraphics[width=0.8\linewidth]{figures/NO_energy.png}
%     \caption{Energies of $NO^+$ molecule at dissociation computed with CASSCF(6, 6), PNOF5 wavefunction, and cumulant that satisfies set of constraints\cite{cumulant_constraints}}
% \end{figure}

\subsection{Spin-Orbital to Spatial 2-RDM Mapping for Closed-Shell Singlet States}
In this section, we detail the mathematical procedure for mapping the spatial two-particle reduced density matrix (2-RDM) to the full spin-orbital 2-RDM for a singlet state.

We define the elements of the spin-orbital 2-RDM using the standard second-quantization expectation value:

$$
 D_{i j k l}^{\sigma_1 \sigma_2 \sigma_3 \sigma_4}=\langle\Psi| a_{i \sigma_1}^{\dagger} a_{j \sigma_2}^{\dagger} a_{l \sigma_4} a_{k \sigma_3}|\Psi\rangle
$$

where $i, j, k, l$ denote spatial orbitals and $\sigma \in\{\alpha, \beta\}$ denotes the spin component.

We define the spatial 2-RDM, $y_{i j k l}$, as the trace over the spin degrees of freedom. For convenience in the derivation, we also define the mixed-spin block $A_{i j k l}$ :

$$
A_{i j k l} \equiv  D_{i j k l}^{\alpha \beta \alpha \beta}=\langle\Psi| a_{i \alpha}^{\dagger} a_{j \beta}^{\dagger} a_{l \beta} a_{k \alpha}|\Psi\rangle
$$

To construct the full spin-orbital tensor $\mathbf{D}$ from the spatial tensor $\mathbf{y}$, one must determine the mixed-spin tensor $\mathbf{A}$. By exploiting the $S U(2)$ spin symmetries of a singlet state, $\mathbf{A}$ can be exactly recovered from the spatial 2-RDM , $y_{ijkl}$, via the following relation:

$$
A_{i j k l}=\frac{2 y_{i j k l}+y_{i j l k}}{6}
$$

Once $A_{i j k l}$ is computed, all six non-zero spin blocks of the spin-orbital 2-RDM (including the off-diagonal spin-flip terms) can be populated purely through permutations of $\mathbf{A}$, as detailed below.

We derive the relation $y_{i j k l}=4 A_{i j k l}-2 A_{i j l k}$ explicitly by applying the total spin-raising operator, $S_{+}=\sum_p a_{p \alpha}^{\dagger} a_{p \beta}$.

Step 1: Expand the spatial trace.
Tracing over the spin components yields the four spin-conserving diagonal blocks:

$$
y_{i j k l}={D_{i j k l}^{\alpha \alpha \alpha \alpha}}+ D_{i j k l}^{\alpha \beta \alpha \beta}+ D_{i j k l}^{\beta \alpha \beta \alpha}+ D_{i j k l}^{\beta \beta \beta \beta}
$$

For a singlet state ( $S_z=0$ ), the state is invariant to global spin inversion. Thus, the pure-spin blocks are equivalent $\left( D_{i j k l}^{\alpha \alpha \alpha \alpha}= D_{i j k l}^{\beta \beta \beta \beta}\right)$ and the mixed-spin blocks are equivalent $\left( D_{i j k l}^{\alpha \beta \alpha \beta}=\right.  D_{i j k l}^{\beta \alpha \beta \alpha}=A_{i j k l}$ ). Substituting these symmetries yields:

$$
y_{i j k l}=2 {D_{i j k l}}^{\alpha \alpha \alpha\alpha}+2 A_{i j k l}
$$

Step 2: Relate the pure-spin and mixed-spin blocks via the $S_{+}$operator.
Because the reference state is a singlet, it is annihilated by the spin-raising operator: $S_{+}|\Psi\rangle=0$. Consequently, the expectation value of the commutator between $S_{+}$and any operator string must be zero. We choose the operator string corresponding to a pure-spin excitation with one mismatched $\beta$ operator:

$$
0=\langle\Psi|\left[S_{+}, a_{i \alpha}^{\dagger} a_{j \beta}^{\dagger} a_{l \alpha} a_{k \alpha}\right]|\Psi\rangle
$$

We expand the commutator by allowing $S_{+}$to act sequentially on each creation and annihilation operator, using the fundamental commutation relations $\left[S_{+}, a_{p \alpha}^{\dagger}\right]=0$, $\left[S_{+}, a_{p \beta}^{\dagger}\right]=a_{p \alpha^{\prime}}^{\dagger}\left[S_{+}, a_{p \alpha}\right]=-a_{p \beta}$, and $\left[S_{+}, a_{p \beta}\right]=0$ :

$$
\left[S_{+}, a_{i \alpha}^{\dagger} a_{j \beta}^{\dagger} a_{l \alpha} a_{k \alpha}\right]=a_{i \alpha}^{\dagger}\left(a_{j \alpha}^{\dagger}\right) a_{l \alpha} a_{k \alpha}+a_{i \alpha}^{\dagger} a_{j \beta}^{\dagger}\left(-a_{l \beta}\right) a_{k \alpha}+a_{i \alpha}^{\dagger} a_{j \beta}^{\dagger} a_{l \alpha}\left(-a_{k \beta}\right)
$$

Taking the expectation value of this expanded sum yields:

$$
0=\left\langle a_{i \alpha}^{\dagger} a_{j \alpha}^{\dagger} a_{l \alpha} a_{k \alpha}\right\rangle-\left\langle a_{i \alpha}^{\dagger} a_{j \beta}^{\dagger} a_{l \beta} a_{k \alpha}\right\rangle-\left\langle a_{i \alpha}^{\dagger} a_{j \beta}^{\dagger} a_{l \alpha} a_{k \beta}\right\rangle
$$

Step 3: Map expectation values to density matrix elements.
We now map the three terms back to our 2-RDM notation:
- Term 1: $\left\langle a_{i \alpha}^{\dagger} a_{j \alpha}^{\dagger} a_{l \alpha} a_{k \alpha}\right\rangle= D_{i j k l}^{\alpha \alpha \alpha}$
- Term 2: $\left\langle a_{i \alpha}^{\dagger} a_{j \beta}^{\dagger} a_{l \beta} a_{k \alpha}\right\rangle= D_{i j k l}^{\alpha \beta \alpha \beta}=A_{i j k l}$
- Term 3: $\left\langle a_{i \alpha}^{\dagger} a_{j \beta}^{\dagger} a_{l \alpha} a_{k \beta}\right\rangle$. To match the standard 2-RDM index ordering, we apply fermion anticommutation to the annihilation operators ( $a_{l \alpha} a_{k \beta}=-a_{k \beta} a_{l \alpha}$ ). This introduces a negative sign, giving $+\left\langle a_{i \alpha}^{\dagger} a_{j \beta}^{\dagger} a_{k \beta} a_{l \alpha}\right\rangle=+A_{i j l k}$.

Substituting these back into the zero-expectation equation:

$$
\begin{gathered}
0= D_{i j k l}^{\alpha \alpha \alpha \alpha}-A_{i j k l}+A_{i j l k} \\
 D_{i j k l}^{\alpha \alpha \alpha \alpha}=A_{i j k l}-A_{i j l k}
\end{gathered}
$$

Step 4: Final Substitution.
We substitute this expression for the pure-spin block into the spatial trace equation from Step 1:

$$
\begin{gathered}
y_{i j k l}=2\left(A_{i j k l}-A_{i j k l}\right)+2 A_{i j k l} \\
y_{i j k l}=4 A_{i j k l}-2 A_{i j l k}
\end{gathered}
$$

This result allows for the exact inversion of the spatial 2-RDM to obtain $A_{i j k l}$.

\subsection{Final Formulas for the Spin-Orbital Blocks}

Using the definitions above and the fundamental anticommutation properties of fermions, the full $2 N \times 2 N \times 2 N \times 2 N$ spin-orbital 2-RDM for a closed-shell singlet state contains exactly six non-zero spin blocks. Expressed entirely in terms of the mixed-spin tensor $A_{i j k l}$, they are:

Same-Spin (Pure) Blocks:

$$
\begin{aligned}
&  D_{i j k l}^{\alpha \alpha \alpha \alpha}=A_{i j k l}-A_{i j l k} \\
&  D_{i j k l}^{\beta \beta \beta \beta}=A_{i j k l}-A_{i j l k}
\end{aligned}
$$

Mixed-Spin (Conserving) Blocks:

$$
\begin{aligned}
&  D_{i j k l}^{\alpha \beta \alpha \beta}=A_{i j k l} \\
&  D_{i j k l}^{\beta \alpha \beta \alpha}=A_{i j k l}
\end{aligned}
$$

Spin-Flip (Off-Diagonal) Blocks:

$$
\begin{aligned}
&  D_{i j k l}^{\alpha \beta \beta \alpha}=-A_{i j l k} \\
&  D_{i j k l}^{\beta \alpha \alpha \beta}=-A_{i j l k}
\end{aligned}
$$

\subsection{Figures}
\setcounter{figure}{0}
\begin{figure}[H] % requires \usepackage{float}
    \centering
    \includegraphics[width=0.8\linewidth]{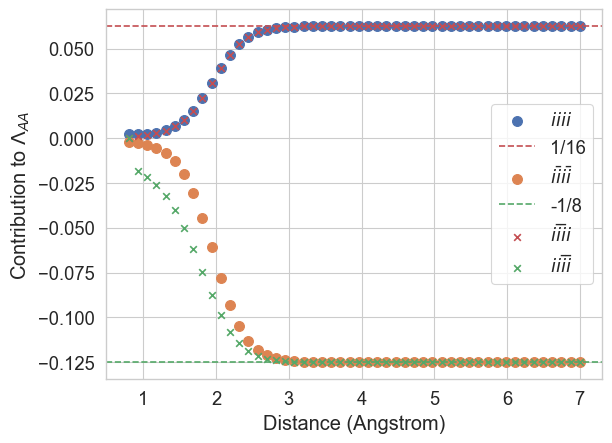}
    \caption{Plot of cumulant matrix elements of sulfur dimer during dissociation.}
\end{figure}

\begin{figure}[H]
    \centering
    \includegraphics[width=0.8\linewidth]{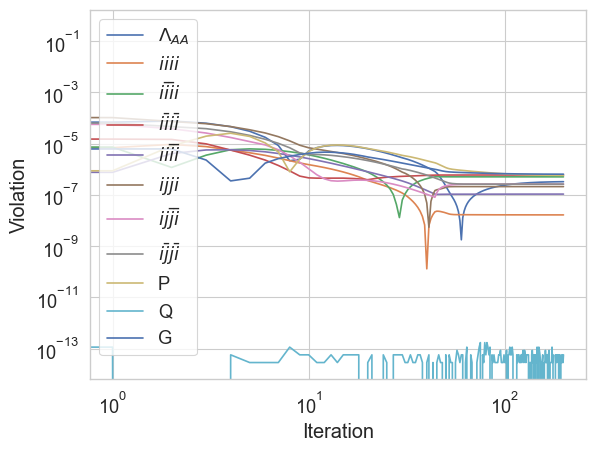}
    \caption{Figure showing how the cumulant constrains are satisfied as a function of iteration.}
\end{figure}

\begin{table}[h!]
\centering
\begin{tabular}{|l|S|S|S|S|S|S|}
\hline
 & {${u}_A$}
 & {${\Lambda}_{AA}$}
 & {$\Lambda_{AA}'$}
 & {$\Lambda_{AB}$}
 & {$\langle \widehat{S}^2 \rangle_A$}
 & {DI} \\
\hline
CASSCF
 & 2
 & -0.5
 & 1.0
 & 0.
 & 2.0
 & 0. \\
\hline
Fixed PNOF5
 & 1.99999138
 & -0.500000000005
 & 1.00001163
 & 0.00000234
 & 2.00000517
 & 0.00001555 \\
\hline
PNOF5
 & 1.99998633
 & -0.49999786
 & 0.49999698
 & 0.00000146
 & 1.49998886
 & 0.00001467 \\
\hline
\end{tabular}
\caption{Table of cumulant violations for PNOF5 versus CASSCF, detailing the asymptotic values of $\left\langle\widehat{S}^2\right\rangle_A$, compacted forms, and the delocalization index (DI) upon dissociation of the \ce{S2} molecule.}
\end{table}

\begin{figure}[H]
    \centering
    \includegraphics[width=0.8\linewidth]{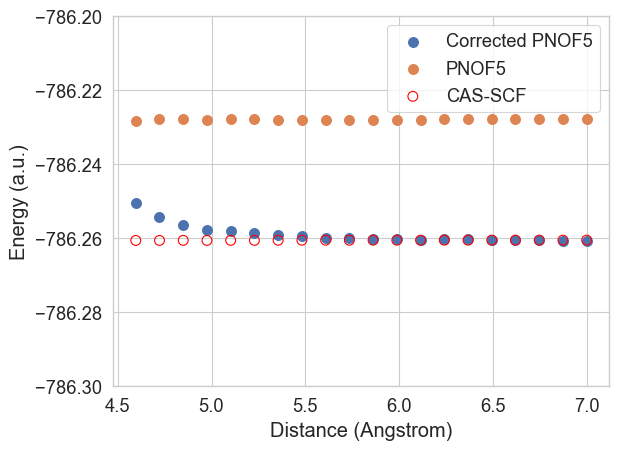}
    \caption{Dissociation energy for the sulfur dimer, with and without corrections, computed using CASSCF(4, 4), the PNOF5 wavefunction, and a cumulant satisfying a set of constraints \cite{cumulant_constraints}.}
\end{figure}

\begin{figure}[H] % requires \usepackage{float}
    \centering
    \includegraphics[width=0.8\linewidth]{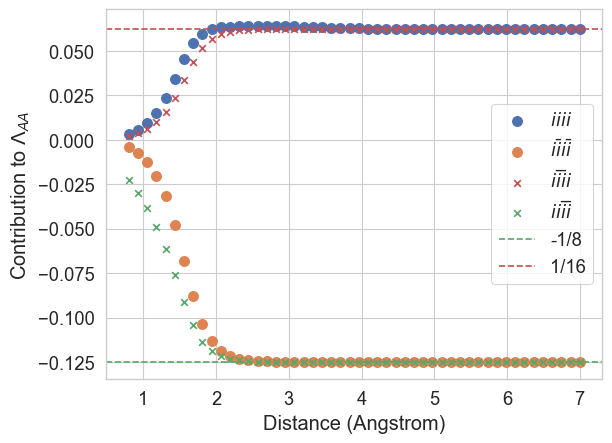}
    \caption{Figure showing how the cumulant contributions in carbon monoxide change with bond length.}
\end{figure}

\begin{table}[h!]
\centering
\begin{tabular}{|l|S|S|S|S|S|S|}
\hline
 & {${u}_A$}
 & {${\Lambda}_{AA}$}
 & {$\Lambda_{AA}'$}
 & {$\Lambda_{AB}$}
 & {$\langle \widehat{S}^2 \rangle_A$}
 & {DI} \\
\hline
CASSCF
 & 3
 & -0.75
 & 2.25
 & 0.
 & 3.75
 & 0. \\
\hline
Fixed PNOF5
 & 2.99999719
 & -0.750000000001
 & 2.25000043
 & 0.00000070
 & 3.74999831
 & 0.00000501 \\
\hline
PNOF5
 & 2.99999592
 & -0.74999893
 & 0.74999892
 & -0.00000006
 & 2.24999694
 & 0.00000403 \\
\hline
\end{tabular}
\caption{Table showing performance of PNOF5 and its correction for cumulant contributions, in the dissociation limit.}
\end{table}

\begin{figure}[H]
    \centering
    \includegraphics[width=0.8\linewidth]{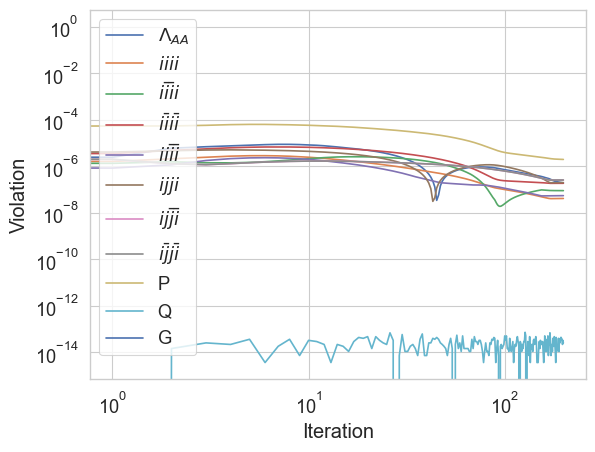}
    \caption{Evolution of constraint violations for the $CO$ molecule in the dissociation regime, showing how constraints become increasingly well satisfied as a function of the number of iterations in the projection algorithm.}
    \end{figure}

\begin{figure}[H]
    \centering
    \includegraphics[width=0.8\linewidth]{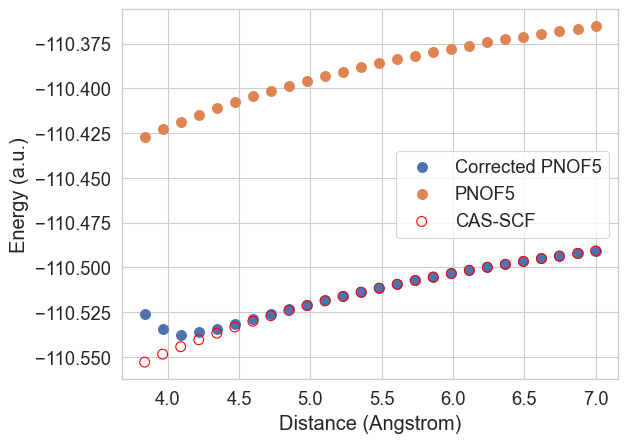}
    \caption{Dissociation energy curve for Carbon monoxide, showing how imposing constraints improves accuracy}
\end{figure}

\begin{table}[h!]
\centering
\begin{tabular*}{\linewidth}{|@{\extracolsep{\fill}}|l|S|S|S|}
\hline
 & {$E, \text{Ha}$}
 & {$\Lambda_{AA}'$}
 & {$\langle \widehat{S}^2 \rangle_A$}\\
\hline
CAS-SCF(6;6)
& -108.7768315
& 2.250000
& 3.749999\\
\hline
Corrected PNOF5
 & -108.7768394
 & 2.24999999
 & 3.74999998\\
\hline
PNOF7
 & -108.7768284
 & 0.74999999
 & 2.24999996\\
\hline
GNOF
 & -108.750156
 & 0.74997537
 & 2.24997792\\
 \hline
 PNOF5
 & -108.6704309
 & 0.74999782
 & 2.24999987\\
\hline
\end{tabular*}
\caption{Asymptotic values of the fragment spins and cumulant contributions in the dissociation limit for the \ce{N2} for the cc-pVDZ basis set.}
\end{table}

\begin{table}[h!]
\centering
\begin{tabular*}{\linewidth}{|@{\extracolsep{\fill}}|l|S|S|S|}
\hline
 & {$E, \text{Ha}$}
 & {$\Lambda_{AA}'$}
 & {$\langle \widehat{S}^2 \rangle_A$}\\
\hline
CAS-SCF(6;6)
& -108.794719
& 2.250001
& 3.7499999\\
\hline
Corrected PNOF5
 & -108.7947184
 & 2.24999997
 & 3.74999999\\
\hline
PNOF7
 & -108.794715
 & 0.74999999
 & 2.24999996\\
\hline
GNOF
 & -108.768245
 & 0.74997537
 & 2.24997792\\
 \hline
 PNOF5
 & -108.6896148
 & 0.74999782
 & 2.24999987\\
\hline
\end{tabular*}
\caption{Asymptotic values of the fragment spins and cumulant contributions in the dissociation limit for the \ce{N2} for the cc-pVTZ basis set.}
\end{table}